\documentclass[11pt]{article}

\usepackage[margin=.75in]{geometry}

\usepackage{amsmath,amssymb}
\usepackage{cancel}
\usepackage{graphicx}
\usepackage{caption}
\usepackage{comment}
\usepackage{subcaption}
\usepackage{color}
\definecolor{indigo}{RGB}{0,0,120}
\usepackage[colorlinks=true, linkcolor=indigo, citecolor=blue, urlcolor=indigo]{hyperref}
\usepackage{mdframed}

\def\tr{\,{\rm tr}\,}

\def\imply{\Rightarrow}

\def\fl{\noindent}

\newcommand{\tl}[1]{\tilde{#1}}
\newcommand{\dd}[2]{\frac {\partial #1}{\partial #2}}

\newcommand{\pdr}{\partial}
\newcommand{\DD}[2]{\frac {d #1}{d #2}}

\newcommand{\beq}{\begin{equation}}
\newcommand{\eeq}{\end{equation}}
\newcommand{\beqs}{\begin{eqnarray}}
\newcommand{\eeqs}{\end{eqnarray}}
\newcommand{\half}{\frac{1}{2}}

\def\al{\alpha} 		
\def\del{\delta}

\def\eps{\epsilon} 
\def\la{\lambda}
		
\def\sig{\sigma}
\def\tht{\theta}
\def\om{\omega}		
\def\Om{\Omega}

\newcommand{\bfx}{{\bf x}}

\newcount\colveccount  
\newcommand*\colvec[1]{\global\colveccount#1  \begin{pmatrix} \colvecnext} \def\colvecnext#1{#1 \global\advance\colveccount-1
        \ifnum\colveccount>0 \\ \expandafter\colvecnext
        \else \end{pmatrix} \fi}

\usepackage{calligra} 
\DeclareMathAlphabet{\mathcalligra}{T1}{calligra}{m}{n}
\DeclareFontShape{T1}{calligra}{m}{n}{<->s*[2.2]callig15}{}

\begin{document}

\title{An introduction to Lax pairs and the zero curvature representation
\author{{\sc Govind S. Krishnaswami and T. R. Vishnu} 
\\ \small
Chennai Mathematical Institute, SIPCOT IT Park, Siruseri 603103, India
\\ \small
Email: {\tt govind@cmi.ac.in and vishnu@cmi.ac.in}}
\date {11 April, 2020\\ \vspace{.25cm}
To appear in the journal \href{https://www.springer.com/journal/12045}{Resonance} published by the Indian Academy of Sciences}}

\maketitle


\abstract{Lax pairs are a useful tool in finding conserved quantities of some dynamical systems. In this expository article, we give a motivated introduction to the idea of a Lax pair of matrices $(L,A)$, first for mechanical systems such as the linear harmonic oscillator, Toda chain, Eulerian rigid body and the Rajeev-Ranken model. This is then extended to Lax operators for one-dimensional field theories such as the linear wave and KdV equations and reformulated as a zero curvature representation via a $(U,V)$ pair which is illustrated using the nonlinear Schr\"odinger equation. The key idea is that of realizing a (possibly) nonlinear evolution equation as a compatibility condition between a pair of linear equations. The latter could be an eigenvalue problem for the Lax operator $L$ and a linear evolution equation generated by $A$, for the corresponding eigenfunction. Alternatively, they could be the first order linear system stating the covariant constancy of an arbitrary vector with respect to the 1+1 dimensional gauge potential $(V,U)$. The compatibility conditions are then either the Lax equation $\dot L = [L, A]$ or the flatness condition $U_t - V_x + [U, V] = 0$ for the corresponding gauge potential. The conserved quantities then follow from the isospectrality of the Lax and monodromy matrices.}

\vspace{.2cm}

{\fl \bf Keywords:} Conserved quantities, Lax pair, Isospectral evolution, Zero curvature representation, Monodromy matrix, Toda chain, Euler top, Rajeev-Ranken model, KdV equation. 

{\scriptsize \tableofcontents}

\normalsize

\section{Introduction}
\label{s:Introduction}

Lax pairs\cite{Lax}, introduced in 1968 by Peter Lax\footnote{Peter David Lax is an American mathematician of Hungarian origin (born 1 May, 1926). He has worked at the Courant Institute of Mathematical Sciences (New York) on various topics including  integrable systems, fluid mechanics and partial differential equations. He received the 2005 Abel prize ``for his groundbreaking contributions to the theory and application of partial differential equations and to the computation of their solutions''.} are a tool for finding conserved quantities of some evolutionary differential equations. A system with possibly nonlinear equations of motion will be said to admit a Lax pair if one can find a pair of matrices/operators $(L,  A)$ such that the equations of motion are equivalent to the Lax equation $\dot L = [L, A]$. If it exists, a Lax pair is not unique. For instance, we may add to $A$ a matrix that commutes with $L$ and add to $L$ a time-independent matrix that commutes with $A$ without altering the Lax equation.  As we will explain, Lax pairs are based on the idea of expressing (typically) nonlinear evolution equations as `compatibility' conditions for a pair of auxiliary {\it linear} equations to admit simultaneous solutions.  The idea only works for certain special systems (integrable systems), which however, play an important role in our understanding of more general dynamical systems. Unfortunately, there is no recipe for finding a Lax pair for a given system or to know in advance whether one exists. So some knowledge of the nature of the system and its solutions (from numerical,  analytical or experimental investigations) coupled with educated guesswork is involved. However, as we will see, once a Lax pair is known, it can be very helpful in understanding the system.

In this expository article, we attempt to give an elementary introduction to the idea of a Lax pair and the associated zero curvature representation. We begin with mechanical systems with finitely many degrees of freedom where the Lax matrices are finite dimensional. A key step is to write the equations in Lax form $\dot L = [L, A]$, where $L$ and $A$ suitable square matrices whose entries depend on the dynamical variables. This makes it easy to read off conserved quantities. Indeed, the Lax equation implies that the eigenvalues of $L$ are independent of time (isospectral evolution). We do this for the harmonic oscillator, Toda chain, Eulerian rigid body and the Rajeev-Ranken model. The latter two examples illustrate how allowing the Lax matrices to depend on an arbitrary spectral parameter gives the method additional power to find conserved quantities. As is well known, each independent conserved quantity imposes one relation among the phase space  variables thereby confining trajectories to a hypersurface of codimension one in the phase space. 

We then generalize the Lax pair framework to continuum systems in one spatial dimension such as the linear wave and KdV equations. $L$ and $A$ are now differential operators, allowing for the possibility of infinitely many conserved quantities. We show how these are obtained for the linear wave and KdV equations. While $A$ may be chosen as first- and third-order operators, it turns out that both equations admit a common Lax operator $L$ (the Schr\"odinger operator) as well as a common infinite tower of conserved quantities. What is more, these quantities are conserved for {\it all} the equations in the so-called KdV hierarchy which is obtained by choosing suitable higher order differential operators for $A$.

We then use the KdV example to pass to a more symmetrical formulation of the Lax pair idea where the Lax equation is viewed as a compatibility condition for a pair of linear equations involving only first derivatives with respect to space and time. This compatibility condition has a geometric meaning: it says that a certain finite dimensional space is flat i.e., a curvature or finite dimensional nonabelian field strength matrix vanishes for all values of a spectral parameter. More generally, we will say that a (nonlinear) system of field equations admits a zero curvature representation if the equations are equivalent to the condition for a certain curvature to vanish. Remarkably, a number of interesting nonlinear field equations especially in one spatial dimension (such as the mKdV, nonlinear Schr\"odinger (see \S \ref{s:NLSE}), sine-Gordon, Heisenberg magnetic chain and principal chiral equations) admit zero curvature representations. What is more, one can use the vanishing of this curvature to obtain (infinitely many) conserved quantities for these systems. This proceeds via the monodromy or parallel transport matrix which governs how vectors change when one goes around a closed spatial loop (for systems with periodic boundary conditions). Our exploration of the Lax pair idea will also come around a full circle when we show that the monodromy matrix itself satisfies a Lax-like equation so that its eigenvalues furnish a set of conserved quantities! Although the monodromy matrix is finite dimensional, one may obtain infinitely many conserved quantities by ensuring that the flatness condition holds for all values of the spectral parameter. The presence of infinitely many conserved quantities gives these systems very remarkable features: aside from admitting exact solutions, these systems typically admit special types of spatially localized solitary waves called solitons which can scatter in complicated ways and yet reemerge while retaining their shapes.

\section{Lax pairs for mechanical systems}
\label{s:Lax-pairs-for-mechanical-systems}

\subsection{Lax pair for the linear harmonic oscillator}
\label{s:SHO}

In the linear harmonic oscillator the displacement $x(t)$ of a particle of mass $m$ from equilibrium evolves according to:
	\beq
	m \ddot x = -k x, \quad \text{where} \quad k > 0 \quad \text{is the force constant}.
	\label{e:EOM-HO}
	\eeq
We will use this example to provide an illustration of the idea of a Lax pair. Though the general solution $x(t) = A \cos(\om t + \phi)$ for constants of integration $A$ and $\phi$ with $\om = \sqrt{k/m}$ is well-known, we will not need the explicit solution to discuss a Lax pair formulation. Introducing the momentum $p = m \dot x$, we may rewrite (\ref{e:EOM-HO}) as a pair of first order equations $\dot x = p/m$ and $\dot p = -m \om^2 x$. It is convenient to regard them as equations for the variables $\om x$ and $p/m$ which have the same dimension (of velocity):
	\beq
	\DD{(\om x)}{t} = \om \left( \frac{p}{m} \right) \quad \text{and} \quad
	\DD{(p/m)}{t} = -\om (\om x).
	\label{e:EOM-HO-dimensional}
	\eeq 
These equations are equivalent to the {\it Lax equation} $\dot L = [L, A]$ for the pair of $2 \times 2$ matrices \cite{B-B-T}
	\beq 
	L = \colvec{2}{p/m & \omega x}{\omega x & -p/m} \quad \text{and} \quad 
	A = \colvec{2}{ 0 & \omega / 2}{-\omega / 2 & 0},
	\label{e:Lax-pair-SHO}
	\eeq
whose entries depend on the dynamical variables $\om x$ and $p/m$. How did we arrive at this $(L, A)$ pair? We notice that (\ref{e:EOM-HO-dimensional}) are linear in $\om \dot x$ and $\dot p/m$. So, for $\dot L = [L, A]$ to reproduce (\ref{e:EOM-HO-dimensional}) we choose $L$ to be linear in $\om x$ and $p/m$. The simplest possibility is to take $L$ to be a $2 \times 2$ real matrix. However, in general this would lead to four EOM. To ensure that there are only 2 independent equations as in (\ref{e:EOM-HO-dimensional}), we will suppose that $L$ is a traceless symmetric matrix with entries linear in $p/m$ and $\om x$ as in (\ref{e:Lax-pair-SHO}). As a consequence, $\dot L$ is also symmetric. Since the commutator of symmetric and anti-symmetric matrices is symmetric, it is natural to take $A$ to be anti-symmetric\footnote{For $[ L, A]$ to be symmetric, $A$ can differ from an anti-symmetric matrix at most by a multiple of the identity, which would not affect the commutator.}. Since the RHS of (\ref{e:EOM-HO-dimensional}) are linear in $\om x$ and $p/m$, we take $A$ to be independent of these variables so that $[L,A]$ would also be linear in them. In terms of these variables, the RHS of (\ref{e:EOM-HO-dimensional}) is independent of $m$ and linear in $\om$, so the entries of the antisymmetric matrix $A$ can depend only on $\om$ and must be linear in it. This essentially leads to the $A$ appearing in (\ref{e:Lax-pair-SHO}). One then verifies that the four Lax equations following from (\ref{e:Lax-pair-SHO}) coincide with (\ref{e:EOM-HO-dimensional}): 
	\beq
	\dot L = \colvec{2}{\dot p/m & \omega \dot x}{\omega \dot x & -\dot p/m} = [L, A] = \colvec{2}{-\om^2 x & \om p/m}{\om p/m & \om^2 x}.
	\eeq
We notice the following feature of the Lax matrix, $\tr L^2 = 2(p^2/m^2 + \om^2 x^2)$ is $(m/4) \times$ the conserved energy of the harmonic oscillator. It turns out that this is a general feature: one may use the Lax matrix to obtain conserved quantities.

\subsection{Isospectral evolution of the Lax matrix} 
\label{s:Isospectral-evolution-of-the-Lax-matrix}

The Lax equation ensures that the eigenvalues (spectrum) of $L$ are independent of time. This property is known as isospectrality. To understand this, let us consider the Lax equation
	\beq
	L_t \equiv \dot{L} = [L, A],
	\label{e:Lax-equation}
	\eeq
where $L$ and $A$ are matrices with entries depending on the dynamical variables. We have used subscripts to denote derivatives. Since the trace of the commutator of a pair of finite dimensional matrices vanishes, $\tr L$ is independent of time\footnote{The Lax equation $\dot L = [L, A]$ bears a resemblance to the Heisenberg equation of motion for an operator $Q$ in the Heisenberg picture of quantum mechanics:
	\beq
	i \hbar\DD{Q}{t} = [Q, H],
	\eeq
where $H$ is the Hamiltonian. If $H$ and $Q$ are finite dimensional matrices, then $\tr [Q, H] = 0$ so that $\tr Q$ is conserved. While the Lax matrices $L$ and $A$ for mechanical systems are finite dimensional,  observables of quantum systems of particles are typically infinite dimensional and unbounded operators. The trace of the commutator of such operators may not vanish (or even be finite). In such cases, $\tr Q$ may not be a (finite) conserved quantity.}. More generally, one may show that the eigenvalues of $L$ are conserved. To see this, we begin with the eigenvalue problem $L \psi = \la \psi$. Differentiating in time,
	\beq
	L_t \psi + L \psi_t = \lambda_t \psi + \lambda \psi_t.
	\eeq
Upon using the Lax equation (\ref{e:Lax-equation}) this becomes
	\beq
	(LA - AL) \psi + L \psi_t = \lambda_t \psi + \lambda \psi_t.
	\eeq
Utilizing $L \psi = \la \psi$ and rearranging, we get
	\beq
	(L - \lambda) A \psi + (L- \lambda) \psi_t = \lambda_t \psi  \quad \text{or} \quad 
	(L- \lambda) (\psi_t + A \psi ) = \lambda_t \psi.
	\label{e:Lax-eqn-and-eignvalue-problem}
	\eeq 
For the eigenvalue $\la$ to be time-independent ($\la_t = 0$), the LHS must vanish. For this to happen, $\psi_t + A \psi$ must be an eigenvector of $L$ with eigenvalue $\la$. Recall that $\psi$ too is an eigenstate of $L$ with the same eigenvalue. Now, for simplicity, we will assume that the $\la$-eigenspace of $L$ is one-dimensional, which implies that $\psi_t + A \psi$ must be a multiple of $\psi$: 
	\beq
	\psi_t + A \psi = \beta \psi
	\label{e:one-d-eigenspace}
	\eeq
for some (possibly time-dependent) complex number $\beta$. This equation may be viewed as an evolution equation\footnote{We may absorb the $\beta {\bf 1}$ term into $A$ (and write $\psi_t = - A \psi$ ) since it commutes with $L$ and therefore does not affect the Lax equation.} for $\psi$:
	\beq
	\psi_t =( -A + \beta {\bf 1}) \psi.
	\eeq
Here, ${\bf 1}$ is the identity matrix. Thus, the Lax equation $L_t = [L, A]$ and this evolution equation for $\psi$ together imply that the eigenvalue $\la$ is a conserved quantity. We say that $L$ evolves isospectrally. 

\vspace{.5cm}

{\fl \bf Alternate demonstrations of isospectrality:} There are ways to show the isospectrality of the Lax matrix $L$ without assuming its eigenspaces are one-dimensional. We give two of them below.
\begin{enumerate}
\item In this approach, we assume that $L$ is hermitian so that $\la$ is real. We take an inner product of $(L- \lambda) (\psi_t + A \psi ) = \lambda_t \psi$ (see Eqn. (\ref{e:Lax-eqn-and-eignvalue-problem})) with the eigenfunction $\psi$ and use hermiticity to get
	\beq
	\langle(L- \lambda)(\psi_{t} + A \psi ),\psi \rangle= \langle \lambda_{t}\psi, \psi \rangle
	\quad \text{or} \quad
	\langle (\psi_{t} + A \psi ), (L- \lambda)\psi \rangle = \la_t \Vert \psi \Vert^2.
	\label{e:nondegenerate}
	\eeq
The LHS vanishes as $L \psi = \la \psi$. Moreover, being an eigenfunction, $||\psi|| \ne 0$, so we must have $\la_t =0$.

\item The isospectrality of $L(t)$ may also be established by showing that $L(t)$ is similar to $L(0)$. Indeed, suppose we define the invertible matrix $S(t)$ via the equation $\dot{S} = -A S$ with the initial condition $S(0) = {\bf 1}$, then the solution of the Lax equation with initial value $L(0)$ is $L(t) = S(t)L(0)S^{-1}(t)$. This is easily verified:
	\beq
	\dot L(t) = \pdr_t (S L(0) S^{-1}) = -ASL(0)S^{-1} - SL(0)S^{-1}\pdr_t(S) S^{-1} = -A L(t) + L(t) A = [L(t), A].
	\eeq
Here we used $\pdr_t(S S^{-1}) = \pdr_t {\bf 1}= 0$, to write $\pdr_t(S^{-1})  = - S^{-1} \pdr_t(S)S^{-1}$. Finally, we observe that two matrices related by a similarity transformation have the same eigenvalues:
	\beq
	L(0) \psi = \la \psi \quad \imply \quad S L(0) S^{-1} (S \psi) = \la (S \psi)
	 \quad \text{or} \quad L(t) (S \psi) = \la (S \psi).
	\eeq
Thus, the eigenvalues of $L$ are conserved in time.

\end{enumerate}
\vspace{.5cm}

{\fl \bf Remark:} In \S \ref{e:From-Lax-pair-to-zero-curvature-representation}, we will revisit this problem from a different viewpoint where the Lax equation $L_t = [L, A]$ is viewed as a compatibility condition among the two linear equations $L \psi = \la \psi$ and $\psi_t = -A \psi$ for constant $\la$.

\subsection{Conserved quantities from the Lax equation}
\label{s:Conserved-quantities-from-Lax-equation}

 We have just seen that if the equations of motion of a system can be written in Lax form $L_t = [L, A]$, then the isospectrality of $L$ gives us conserved quantities. These conserved quantities\footnote{A mechanical system with $p$ degrees of freedom can have at most $2p-1$ independent conserved quantities. So though $\tr L^n$ is conserved for any positive integer $n$, not all of them may be independent.}  could be the eigenvalues of $L$ or equivalently the spectral invariants $\det L$ and $\tr L^n$ for $n = 1, 2, 3, \ldots$. For example, the familiar conserved energy of the harmonic oscillator may be expressed in terms of the Lax matrix of Eqn. (\ref{e:Lax-pair-SHO}):
	\beq
	E = \half \left( \frac{p^2}{m} + m \omega^2 x^2 \right) =  -\frac{m}{2} \det L = \frac{m}{4} \tr L^2.
	\label{e:Energy-SHO}
	\eeq
We also notice that for any $E > 0$ this Lax matrix has two distinct eigenvalues $(\pm \sqrt{2E/m})$ leading to 1D eigenspaces (one linearly independent eigenvector for each eigenvalue), as was assumed in Eqn. (\ref{e:one-d-eigenspace}). Furthermore, for $n = 1, 2, 3, \ldots$,
	\beq
	L^{2n} = \left( \frac{p^2}{m^2} + x^2 \omega^2 \right)^n {\bf 1} = \left( \frac{2 E}{m}\right)^n {\bf 1}
	\quad \text{and} \quad
	L^{2n+1} = \left( \frac{2 E}{m}\right)^n L.
	\eeq
Thus, $\tr L^{2n} = 2 \left( 2E/m \right)^n$ while $\tr L^{2n+1} = 0$ so that the traces of higher powers of $L$ do not furnish any new conserved quantities. Indeed, a system with one degree of freedom cannot have more than one independent conserved quantity. In fact, the conservation of energy restricts the trajectories of the harmonic oscillator to lie on a family of ellipses in the $x$-$p$ phase plane. If there was an additional conserved quantity, trajectories would reduce to points which cannot describe nontrivial time evolution. 

We now discuss some more examples of Lax representations. Our first example is the Toda chain which admits a simple and elegant Lax pair. We then consider the Euler equations for a rigid body. They admit a simple Lax pair, which however does not allow us to obtain its conserved energy. This problem is solved by introducing a new Lax pair with a `spectral parameter'. We will also give a Lax pair with spectral parameter for the equations of the Rajeev-Ranken model, which can be viewed as a generalization of the Euler equations to a centrally extended Euclidean algebra \cite{G-V-2}. 

\subsection{Toda chain: Lax pair and conserved quantities}
\label{s:Toda-chain}
\begin{figure}[h]
	\begin{center}
	\includegraphics[width=8cm]{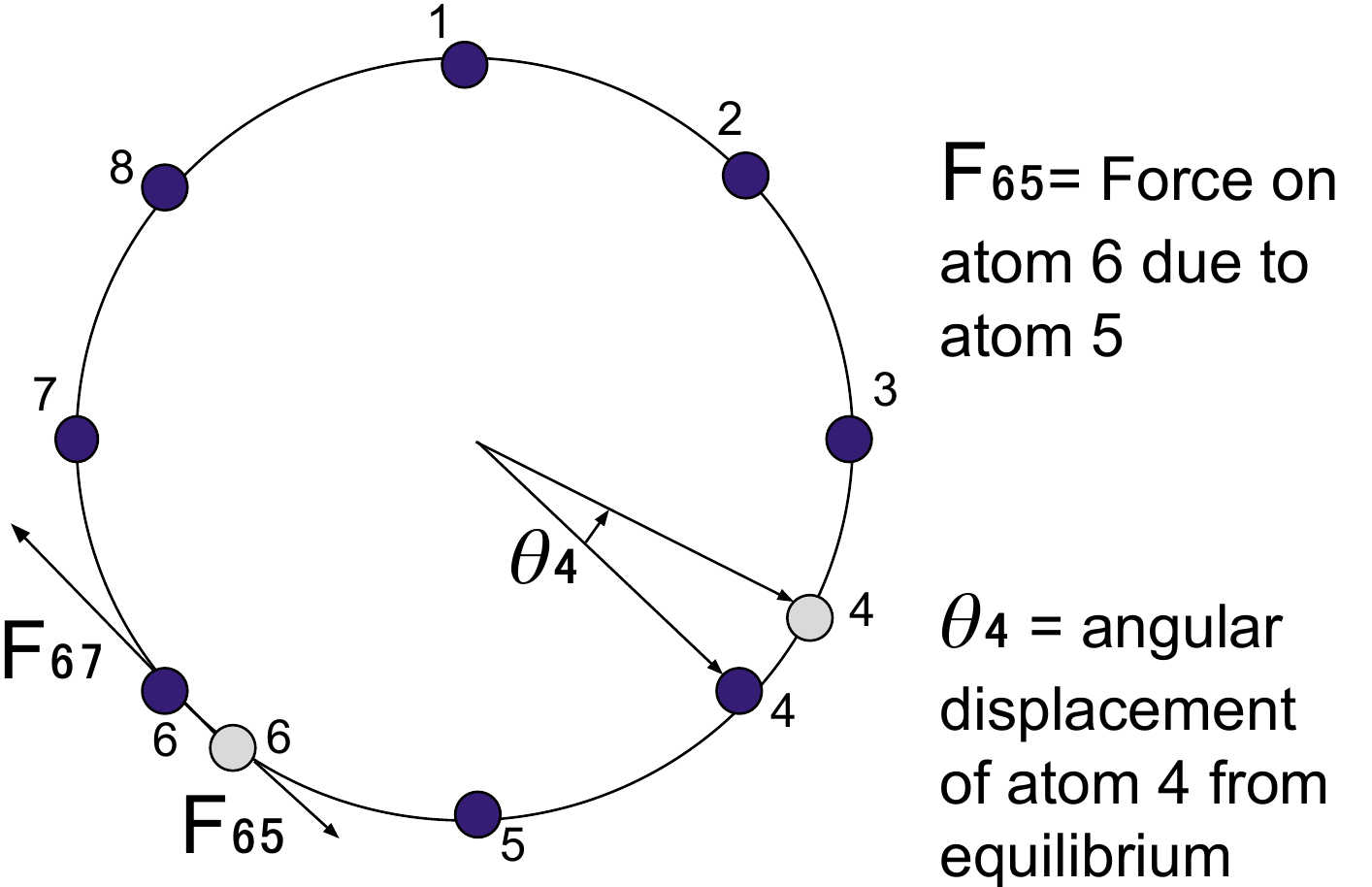}
	\caption{\footnotesize Toda chain of $N = 8$ particles with periodic boundary conditions.}
	\label{f:Toda-chain}
	\end{center}
	\end{figure}

In 1967, Morikazu Toda introduced a model for a one-dimensional crystal in which a chain of identical atoms/particles of mass $m$ interact with their nearest neighbours via nonlinear springs with exponential forces. If $x_i$ is the displacement of the $i^{\rm th}$ particle from its equilibrium position and $p_i$ its momentum, then the EOM are
	\beq
	m \dot{x_i} = p_i \quad \text{and} \quad
	\dot{p_i} = \kappa \: \left( e^{-(x_i - x_{i-1})} - e^{-(x_{i+1} - x_i)} \right).
	\label{e:EOM-Toda}
	\eeq
Here, $\kappa$ is a force constant and we will work in units where $\kappa = m = 1$. We will consider an $N$ particle Toda chain subject to periodic boundary conditions: $x_{N + i} = x_i$ for all $i$. Thus, we may visualize the particles as lying on a circle and interpret $x_i$ as the angular displacement $\tht_i$ from equilibrium (see Fig.~\ref{f:Toda-chain}).
The exponential nonlinearity of the EOM (\ref{e:EOM-Toda}) may be made quadratic by introducing Falschka's variables \cite{Flaschka}
	\beq
	a_i = \half e^{-(x_i - x_{i-1})/2} \quad \text{and} \quad
	b_i = -\half p_{i-1},
	\label{e:Flaschka-Variables}
	\eeq	
which evolve according to
	\beq
	\dot{a_i} = a_i (b_{i+1} - b_i) \quad \text{and} \quad
	\dot{b_i} = 2(a_i^2 - a_{i-1}^2).
	\label{e:EOM-Toda-Flaschka-Variables}
	\eeq
These equations are equivalent to the Lax equation $\dot L = [L, A]$ if we define the essentially tridiagonal matrices $L$ and $A$ as below
	\footnotesize
	\beq
	L = \colvec{5}{b_1 & a_1 & 0 & \cdots & a_N}{a_1 & b_2 & a_2} {0 & a_2 & b_3}{\vdots & & & \ddots}{a_N & & & & b_{N}}
	\;\; \text{and} \;\;
	A = \colvec{5}{0 & -a_1 & 0 &  \cdots & a_N}{a_1 & 0 & -a_2 &  \cdots & 0}{0 & a_2 & 0 &  \cdots & 0}{\vdots & & & & \vdots}{-a_N & & & &  0}. \qquad
	\eeq	
	\normalsize
The traces $\tr L^n$ for $n = 1, 2, \ldots, N$ or alternatively the coefficients of the characteristic polynomial $\det(L - \la {\bf 1})$ give us $N$ conserved quantities \cite{Henon}. The first two of these may be interpreted in terms of the total momentum and energy of the chain
	\beqs
	\tr L &=& \sum_{i =1}^N b_i = -\half \sum_{i = 1}^N p_{i-1} = -\frac{P}{2} \quad \text{and}\cr
	\tr L^2 &=& \sum_{i =1}^N \left( 2 a_i^2 + b_i^2 \right) = \half \sum_{i = 1}^N \left( \half p_i^2 + e^{-(x_i - x_{i-1})} \right) = \frac{E}{2}. \quad
	\eeqs

\subsection{Euler top: Lax pair and conserved quantities}
\label{s:Euler-top}
We consider a rigid body (e.g. a top) free to rotate about its center of mass (which is held fixed) in the absence of external forces like gravity. In a frame that rotates with the body, its EOM may be written as a system of three first order `Euler' equations \cite{L-L, H-F} for the components of angular momentum $\vec S$ about its center of mass:
	\beqs
	\vec S_t &=& \vec S \times \vec \Om
	\quad \text{or} \quad 
	\dot S_1 = S_2 \Om_3 - S_3 \Om_2, \cr 
	\dot{S}_2 &=& S_3 \Om_1 - S_1 \Om_3 \quad \text{and} \quad
	\dot{S}_3 = S_1 \Om_2 - S_2 \Om_1.
	\label{e:Euler-top-EOM}
	\eeqs	
Here, $\vec \Om = (\Om_1, \Om_2, \Om_3)$ is the angular velocity vector which is related to $\vec S = (S_1, S_2, S_3)$ via  $\vec S = I \vec \Om$. The inertia tensor $I_{ij} = \int (\bfx^2 \del_{ij} - x_i x_j) \rho(\bfx) \: d\bfx$ is a $3 \times 3$ real symmetric  matrix which encodes the distribution of the mass density $\rho({\bf x})$. The eigenvalues $I_1, I_2$ and $ I_3$ of $I$ are called the {\it principal moments of inertia}. In what follows, we will choose the axes of the co-rotating frame to be the principal axes of inertia (eigenvectors of $I$) so that the inertia tensor becomes diagonal: $I = {\rm diag}(I_1, I_2, I_3)$. 
	
There is a straightforward way of expressing the Euler equations in Lax form if we introduce the anti-symmetric matrices 
	\beq
	S = \colvec{3}{0 & S_3 & -S_2}{-S_3 & 0 & S_1}{S_2 & -S_1 & 0} \quad \text{and} \quad 
	\Om = \colvec{3}{0 & \Om_3 & -\Om_2}{-\Om_3 & 0 & \Om_1}{\Om_2 & -\Om_1 & 0}
	\eeq 
corresponding to the vectors $\vec S$ and $ \vec{\Om}$. These matrices are obtained via an isomorphism from the $\mathbb{R}^3$ to the $\mathfrak{so}(3)$ Lie algebra by contracting with the Levi-Civita symbol, e.g. $S_{ij} = \sum_k \eps_{ijk} S_k$ while conversely $S_k = \sum_{i,j} \eps_{ijk} S_{ij}/2$. The cross product
	\beq
	\vec S \times \vec \Om = ( S_2 \Om_3 - S_3 \Om_2, S_3 \Om_1 - S_1 \Om_3, S_1 \Om_2 - S_2 \Om_1),
	\eeq
then corresponds to (the negative of) the matrix commutator:
	\small
	\beq
	[S, \Om] = \colvec{3}{0 & S_2 \Om_1 - S_1 \Om_2   & S_3 \Om_1 - S_1 \Om_3}{S_1 \Om_2 - S_2 \Om_1& 0 & S_3 \Om_2 - S_2 \Om_3}{S_1 \Om_3 - S_3 \Om_1 &  S_2 \Om_3 - S_3 \Om_2 & 0}.
	\eeq \normalsize
Thus, the Euler equations (\ref{e:Euler-top-EOM}) take the Lax form:
	\beq
	S_t = [ \Om, S ].
	\label{e:Euler-top-EOM-matrix}
	\eeq
Comparing with (\ref{e:Lax-equation}) we see that $(S,-\Om)$ furnish a Lax pair. What is more, the Lax equation then implies that $-(1/2)\tr S^2 = S_1^2 + S_2^2 + S_3^2$ (square of angular momentum) is a conserved quantity. Indeed, it is straightforward to check using (\ref{e:Euler-top-EOM}) that $S_1 \dot{S}_1 + S_2 \dot{S}_2 + S_3 \dot{S}_3 = 0$. In addition to $\vec S^2$, the Euler top is known to possess another conserved quantity, its energy: 
	\beq
	E = \half \left( \frac{S_1^2}{I_1} + \frac{S_2^2}{I_2} + \frac{S_3^2}{I_3} \right).
	\eeq
However, $E$ depends on the principal moments of inertia and cannot be obtained from the Lax matrix $S$ by combining the traces of its powers as $S$ is independent of $I_{1,2,3}$.

Thus, we seek a new Lax pair $(L,A)$ such that both $\vec S^2$ and $E$ can be obtained from traces of $L$. We therefore introduce a new Lax matrix which is a combination of the angular momentum and inertia matrices, weighted by a parameter\footnote{The parameter $\la$ that appears in the Lax matrix $L$ of Eqn. (\ref{e:Lax-pair-Euler-top}) is (somewhat confusingly) known as a spectral parameter. It is not to be confused with the symbol for an eigenvalue of the Lax matrix! The reason for this terminology will be clarified in \S \ref{e:From-Lax-pair-to-zero-curvature-representation}.}  $\la$. However, in place of $I$, it turns out to be convenient to work with the diagonal matrix: ${\cal I} = {\rm diag}({\cal I}_1, {\cal I}_2, {\cal I}_3)$ with ${\cal I}_k = (1/2)(I_{i} + I_{j} - I_{k})$ where $(i, j, k)$ is any cyclic permutation of $(1, 2, 3)$. For example, ${\cal I}_1 = (1/2)(I_2 + I_3 - I_1)$. Now, we postulate the new Lax pair \cite{B-B-T}
	\small
	\beqs
	L(\la) &=& {\cal I}^2 + \frac{S}{\la} = \colvec{3}{{\cal I}_1^2 & S_3/\la & -S_2/\la}{-S_3/\la & {\cal I}_2^2 & S_1/\la}{S_2/\la & -S_1/\la & {\cal I}_3^2} \quad \text{and} \cr
	A(\la) &=& -(\la {\cal I} + \Om) = -\colvec{3}{ \la {\cal I}_1& \Om_3 & -\Om_2}{-\Om_3 & \la {\cal I}_2 & \Om_1}{\Om_2 & -\Om_1 & \la {\cal I}_3}.
	\label{e:Lax-pair-Euler-top}
	\eeqs 
	\normalsize
To motivate this Lax pair we first note that putting $L = S/\la$ and $A = -\Om$ in $\dot L = [L,A]$ gives the desired EOM (\ref{e:Euler-top-EOM-matrix}). For the energy to emerge as a conserved quantity from $\tr L^2$, we will augment this Lax pair by matrices involving the principal moments of inertia (or the matrix $\cal I$) while ensuring that the EOM are not affected. Since ${\cal I}_{1,2,3}$ are constant in time we can add any matrix function $f({\cal I})$ to $L$ without affecting $\dot L$. However, this will affect the commutator $[L, A]$. To cancel this contribution we will add another matrix function $g({\cal I})$ to $A$. Thus, $L = S/ \la + f$ and $A = - (\Om + g)$. For the unwanted terms $[S/ \la, g]$ and $[f, \Om]$ in $[L, A]$ to have a chance of cancelling, we use the relation $\vec S = I \vec\Om$ and dimensional analysis to pick $f = {\cal I}^2$ and $g = \la {\cal I}$ as in (\ref{e:Lax-pair-Euler-top}). Some algebra now shows that the Lax equation $\dot L = [L, A]$ is equivalent to (\ref{e:Euler-top-EOM-matrix}). Indeed, 
	\beq
	\dot{L} - [L, A] = \frac{1}{\la}(\dot S + [S, \Om]) +  [S, {\cal I}] + [{\cal I}^2, \Om].
	\eeq
Using $\vec S = I \vec \Om$, one finds that the sum $[S, {\cal I}]+[{\cal I}^2, \Om]$ vanishes. Thus, requiring the Lax equation to hold for any value of $\la$ leads to the Euler equations for the angular momentum vector $\vec S$ as in (\ref{e:Euler-top-EOM-matrix}). 

\begin{figure}[h]
	\begin{center}
	\includegraphics[width=7cm]{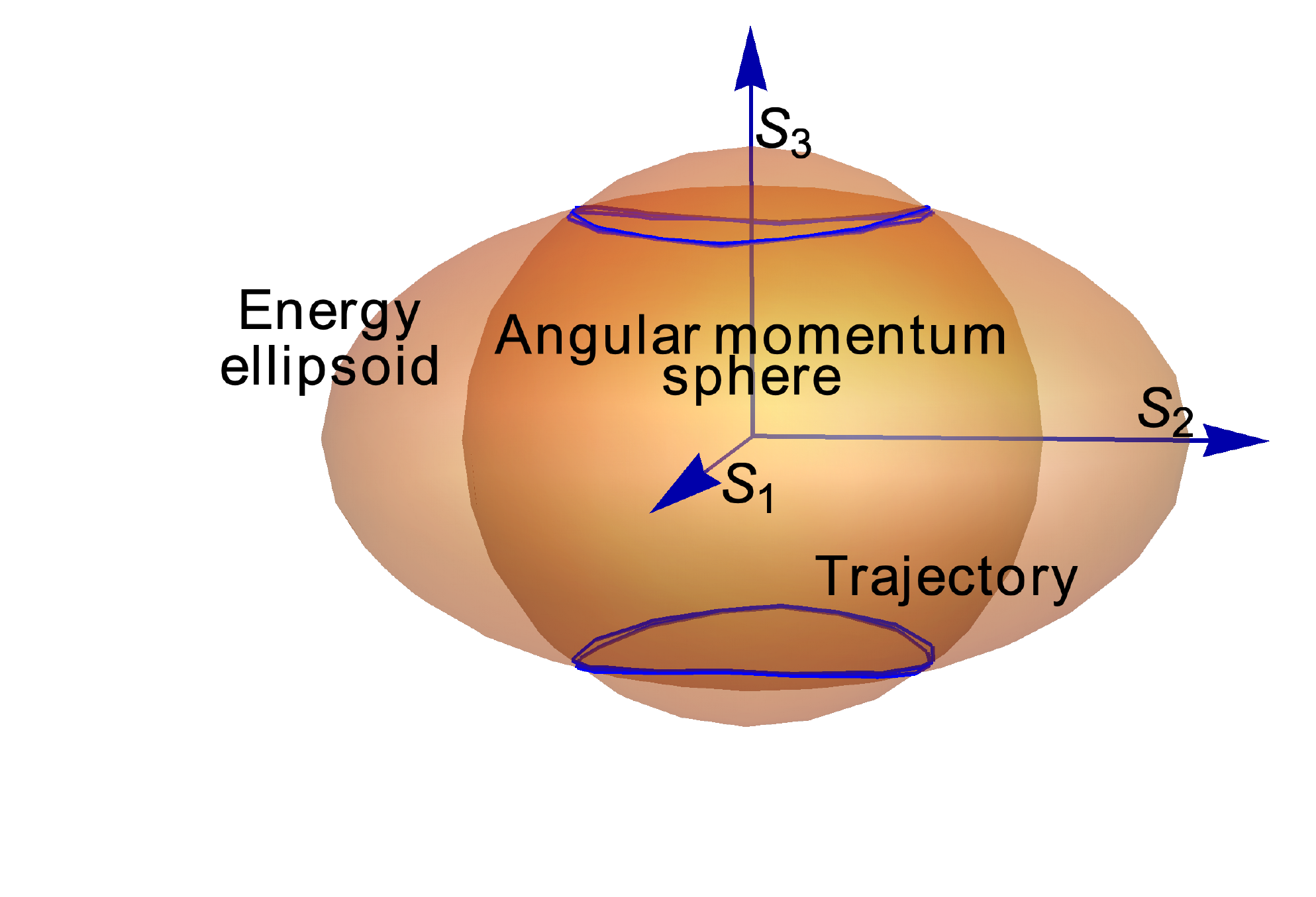}
	\caption{\footnotesize The intersection of the energy ellipsoid and angular momentum sphere is the orbit of the angular momentum vector $\vec S$ in the corotating frame of the Euler top.}
	\label{f:Euler-top-CL-set}
	\end{center}
	\end{figure}

The trace of this new Lax matrix $L$ is conserved, but it is not a dynamical variable as it is simply a quadratic polynomial in the material constants $I_{1,2,3}$. Pleasantly, the traces of the second and third powers\footnote{Traces of higher powers of $L$ also lead to conserved quantities but they are simply functions of $\vec S^2$ and $E$.} of $L$ involve the square of angular momentum $\vec S^2$ and energy $E$, allowing us to deduce that both of them are conserved:
	\beqs
	& \tr L^2 = \tr {\cal I}^4 - \frac{2}{\la^2} \vec{S}^2 \qquad \text{and} \cr
	& \tr L^3 = \text{tr} \left[ {\cal I}^6 + \frac{3}{\la^2} {\cal I}^2 S^2 \right] = \text{tr} \, {\cal I}^6 - \frac{3}{\la^2} \left(\frac{(\text{tr} \: I)^2 \vec{S}^2}{4} - I_1 I_2 I_3 E \right). \qquad
	\eeqs
These conservation laws may be used to determine how $\vec S$ evolves in the corotating frame. Indeed, since both $E$ and $\vec S^2$ are conserved, trajectories must lie along the intersection of the energy ellipsoid and angular momentum sphere:
	\beq
	E = \half \left( \frac{S_1^2}{I_1} + \frac{S_2^2}{I_2} + \frac{S_3^2}{I_3} \right)
	\quad \text{and} \quad
	\vec S \cdot \vec S = S_1^2 + S_2^2 + S_3^2.
	\eeq
These two quadratic surfaces typically intersect along a closed curve which forms the periodic orbit of the tip of the angular momentum vector $\vec S$ as shown in Fig.~\ref{f:Euler-top-CL-set}. 

Having found the evolution of the angular momentum vector, one still needs to use $\vec{S}(t)$ to solve three first order equations for the `Euler angles' ($\tht, \phi$ and $\psi$) to find the instantaneous orientation of the rigid body in space. In the absence of external forces, a top displays two types of motion: spinning about an instantaneous axis of rotation and precession of this axis about the fixed direction of angular momentum in the lab frame. For more on this, see the discussion in \cite{L-L}.

\subsection{Rajeev-Ranken model}
\label{s:RR-model}

\begin{figure}[h]
	\begin{center}
	\includegraphics[width=5cm]{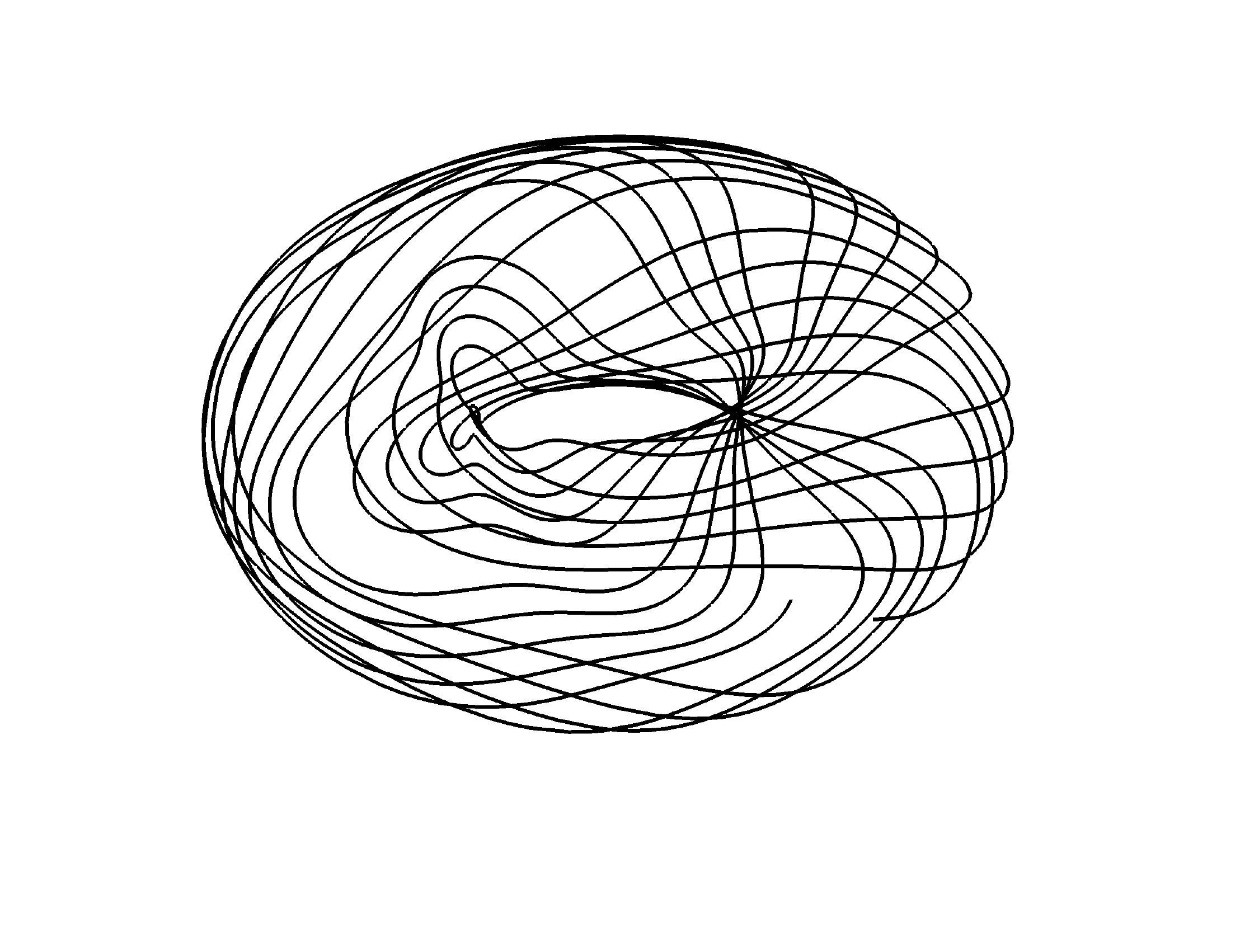}
	\caption{\footnotesize A quasi-periodic trajectory on a torus which is the intersection of level surfaces of the four conserved quantities of the Rajeev-Ranken model.}
	\label{f:theta-phi-dynamics-3D}
	\end{center}
	\end{figure}

The Rajeev-Ranken model \cite{R-R, G-V-1} describes certain nonlinear `continuous waves' in a one-dimensional medium\footnote{The Rajeev-Ranken(RR) model is a mechanical reduction of a 1+1-dimensional scalar field theory (with field equations $\ddot{\phi} = \phi'' + \la [ \dot{\phi}, \phi']$ for the $\mathfrak{su}(2)$ Lie algebra-valued field $\phi(x,t)$) dual to the SU(2) principal chiral model.  It describes the dynamics of nonlinear screw-type waves of the form $\phi(x, t) = e^{K x} R(t) e^{-K x} + mKx$. Here, $K$ is a constant $\mathfrak{su}(2)$ matrix, $m$ a dimensionless parameter and $\la$ a dimensionless coupling constant. The variables of the RR model $J$ and $S$ are related to the anti-hermitian $2 \times 2$ matrix $R$ via $J = [K, R] + m K$ and $S = \dot R + K/ \la$.}. This model is nice for our purposes since it is possible to discover a Lax pair for its equations almost by inspection! Very roughly, it is a generalization of the Euler top with two dynamical 3-vectors $\vec J(t)$ and $\vec S(t)$ and a six-dimensional phase space. Their evolution is governed by the pair of equations
	\beq
	\dot{\vec J} = \vec K \times \vec S \quad \text{and} \quad
	\dot{\vec S} = g \left( \vec S \times \vec J \right).
	\label{e:EOM-JS}
	\eeq
Here, $g$ is a positive constant and $\vec K = - k \hat z$ is a constant vector taken along the $z$-axis, where $k$ is a constant with dimensions of wavenumber. As for the Euler top (see \S \ref{s:Euler-top}), we may rewrite these equations in matrix form by defining the anti-hermitian matrices $J, S$ and $K$: 
	\beq
	J = \vec J \cdot \frac{\vec \sig}{2 i}, \quad  S = \vec S \cdot \frac{\vec \sig}{2 i} \quad \text{and} \quad K = -k \frac{\sig_3}{2 i}.
	\eeq
Here, $\vec \sig$ is the vector whose components are the Pauli matrices $\sig_{1,2,3}$.
The EOM (\ref{e:EOM-JS}) now take the form
	\beq
	\dot{J} = \left[K, S\right] 
	\quad \text{and} \quad 
	\dot{S} = g \left[S, J\right].
	\label{e:EOM-JS-matrix}
	\eeq
They admit a Lax representation if we postulate the Lax pair\footnote{To arrive at this Lax pair we notice that $\dot L = [L, A]$ can lead to (\ref{e:EOM-JS-matrix}) if $J$ and $S$ appear linearly in $L$ as coefficients of different powers of $\la$.}  
 	\beq
	L(\la) = -K \la^2 + J \la + \frac{S}{g} 
	\quad \text{and} \quad	
	A(\la) = -\frac{S}{\la}.
	\label{e:Lax-pair}
	\eeq
Indeed, requiring the Lax equation $\dot L = [L, A]$ to hold for all values of the spectral parameter $\la$ leads to Eqn. (\ref{e:EOM-JS-matrix}) along with the condition $\dot K = 0$, which is consistent with the constancy of the vector $\vec K$. As a consequence of this Lax representation, $\tr L^n(\la)$ must be conserved for any $\la$ and for any $n = 1,2,3 \ldots$. Thus, each coefficient of the $(2n)^{\rm th}$ degree polynomial $\tr L^n(\la)$ furnishes a conserved quantity. For instance, $\tr L \equiv 0$ but 
	\beq
	-2 \tr L^2 = \la^4 \: \vec K^2 - 2 \la^3 \: \vec J \cdot \vec K + 2 \la^2 \left( \frac{\vec J^2}{2} - \frac{\vec S \cdot \vec K}{g} \right)+ \frac{2 \la}{g} \: \vec S \cdot \vec J + \frac{\vec S^2}{g^2} 
	\eeq
leads to four conserved quantities: 
	\beq
	\vec K \cdot \vec J = - k J_3, \quad
	\half \vec J^2 + \frac{k}{g} S_3,\quad 
	\vec S \cdot \vec J
	\quad \text{and} \quad
	\vec S^2.
	\label{e:conserved-quantities}
	\eeq
It turns out that the traces of odd powers of $L$ are identically zero while $\tr L^4, \tr L^6, \ldots$ do not lead to any new conserved quantities. We may interpret the four conserved quantities of (\ref{e:conserved-quantities}) geometrically in the three-dimensional spaces of $\vec S$ and $\vec J$ vectors. For instance, the constancy of $\vec S^2$ implies that $\vec S$ is confined to a sphere. For $\vec S$ lying on such a sphere, the conservation of $\vec S \cdot \vec J$ implies that $\vec J$ must lie on a plane perpendicular to $\vec S$. Similarly, constancy of $J_3$ defines a horizontal plane and that of $\vec J^2/2 + (k/g) S_3$ defines a sphere for each such vector $\vec S$. Remarkably, it turns out that the intersection of these four surfaces is almost always a two-dimensional torus (surface of a vada/doughnut) \cite{G-V-2} in the six-dimensional space of $(\vec S, \vec J)$ pairs. This implies that unlike in the Euler top where $\vec S$ was periodic, here, trajectories are typically quasi-periodic\footnote{Quasi-periodic refers to a superposition of two periodic motions (with incommensurate frequencies) corresponding to the two cycles (`small' and `large' non-contractible loops) of the torus.} and fill up the whole torus as in Fig.~\ref{f:theta-phi-dynamics-3D}. 	

A Lax pair similar to the one in Eqn. (\ref{e:Lax-pair}) applies to the Neumann model \cite{B-B-T, G-V-1} which describes the motion of a particle on a sphere subject to harmonic forces due to springs attached to the coordinate hyperplanes.

\section{Lax pairs and zero curvature representations for wave equations}
\label{s:Lax-pairs-and-zero-curvature-representations-for-wave-equations}

So far, we considered systems of particles with finitely many degrees of freedom. In this section, we extend the idea of a Lax pair to certain continuum mechanical systems with infinitely many degrees of freedom (systems of fields rather than finitely many particles). We will do this in the context of the linear wave equation for vibrations of a stretched string and the nonlinear Korteweg de-Vries (KdV) equation for water waves. The Lax pair framework will also be given a geometric reformulation in terms of the vanishing of a certain curvature, allowing us in principle to find infinitely many conserved quantities for certain field equations. 
 
\subsection{First order linear wave equation}
\label{s:wave-equation}

One of the simplest field equations in one dimension (1D) is the wave equation\footnote{The 1$^{\rm st}$ order wave equation $u_t + c u_x  = 0$ is related to the 2$^{\rm nd}$ order wave equation $(\pdr^2_t - c^2 \pdr^2_x) \phi = 0$. Indeed, the d'Alembert wave operator $\pdr^2_t - c^2 \pdr^2_x$ may be factorized as $(\pdr_t + c \pdr_x )(\pdr_t - c \pdr_x)$. The first order equations $u_t + c u_x = 0$ and $v_t - c v_x = 0$ describe right/left-moving waves $u = f(x - ct)$ and $v = g(x + ct)$ while the 2$^{\rm nd}$ order wave equation describes bi-directional propagation: $\phi(x, t) = f(x - ct) + g(x + ct)$.}:
	\beq
	u_t + c u_x = 0 \quad \text{for constant} \quad c.
 	\label{e:1D-wave-eqn}
	\eeq
Here, $u(x, t)$ could represent the amplitude/height of the wave (sound/water etc.) at position $x$ at time $t$. Here, subscripts on $u$ denote partial derivatives. For $c > 0$, this partial differential equation  describes right-moving waves that travel at speed $c$ while maintaining their shape.
Indeed, one checks that $u(x, t) = f(x - ct)$ is a solution of (\ref{e:1D-wave-eqn}) for any differentiable function $f$. We seek a Lax pair of differential operators $L$ and $A$ (depending on $u$) such that $L_t = [L, A]$ is equivalent to (\ref{e:1D-wave-eqn}). It is convenient to take $L$ to be the Schr\"odinger operator $L  = -\pdr^2 + u(x,t)$, where $\pdr = \pdr_x = \pdr/\pdr x$. $L$ is familiar from Sturm-Liouville theory as well as from quantum mechanics as the Hamiltonian\footnote{In one-dimensional quantum systems, bound state energy eigenvalues are nondegenerate (see p.~99 of \cite{Griffiths}). Thus, the corresponding eigenspaces of $L = -\pdr_x^2 + u$ are one-dimensional. See the discussion in \S \ref{s:Isospectral-evolution-of-the-Lax-matrix}.}  of a particle moving in the potential $u(x,t)$. Since $L$ is symmetric (hermitian), $L_t = u_t$ is also symmetric, so for the Lax equation to make sense $[L, A]$ must also be symmetric. As in \S \ref{s:Isospectral-evolution-of-the-Lax-matrix}, choosing $A$ to be anti-symmetric (up to the addition of an operator that commutes with $L$) guarantees this. It turns out that $A = c \,\pdr$ does the job (see Appendix \ref{a:Finding-a-Lax-operator-$A$-for-the-wave-equation}). Indeed, using the commutator $[\pdr , f] = f'$ for any function $f$, we see that the Lax equation is equivalent to the wave equation:
	\beq
	L_t = u _t = [L, A] = [-\pdr^2 + u(x,t), c \pdr] = [u, c \pdr] = -c u_x.
	\label{e:wave-eqn-Lax-eqn-form}
	\eeq
We will use this Lax pair as a stepping stone to find a Lax pair for the KdV equation which is a nonlinear wave equation with widespread applications. As discussed in \S \ref{s:Lax-pairs-for-mechanical-systems}, the existence of a Lax pair is usually associated with the presence of conserved quantities. For example, integrating (\ref{e:1D-wave-eqn}) in $x$, we get 
	\beq
	\DD{}{t} \int_{-\infty}^{\infty} u \: dx = -c\int_{-\infty}^{\infty}  u_x dx = -c( u(\infty) - u(-\infty)) = 0
	\eeq
assuming $u \to 0$ as $x \to \pm \infty$. Thus, $C_1 = \int_{-\infty}^{\infty} u \: dx$ is conserved. The reason this worked is that (\ref{e:1D-wave-eqn}) takes the form of a local conservation law: $\pdr_t \rho + \pdr_x j = 0$ with $\rho = u$ and $j = c u$. Integrating an equation in local conservation form implies the conservation of $\int_{-\infty}^{\infty} \rho \: dx$, provided the `flux' of $j$ across the `boundary' vanishes: $j(\infty) - j(-\infty) = 0$. Similarly, multiplying (\ref{e:1D-wave-eqn}) by $u$ leads to an equation that is again in local conservation form: $\pdr_t (u^2/2) + c \pdr_x (u^2/2) = 0$. Thus, $C_2 = \int_{-\infty}^{\infty} u^2 \: dx$ is {\it also} conserved. In a similar manner, we find that $\pdr_t u^n + c \pdr_x u^n = 0$, so that $C_n = \int_{-\infty}^{\infty} u(x,t)^n \: dx $ is conserved for any $n = 1,2,3, \ldots$. Thus, the wave equation admits infinitely many constants of motion.


However, unlike in \S \ref{s:Lax-pairs-for-mechanical-systems}, $C_n$ have not been obtained from the Lax operator $L$. As we will see in \S \ref{s:Infinitely-many-conserved-quantities-for-the-linear-wave-equation}, the wave equation also admits {\it another} infinite sequence of conserved quantities $Q_n$ that may be obtained from $L$. Unlike $C_n$, the $Q_n$ turn out to be very special: they are conserved quantities both for the wave equation and its upcoming nonlinear generalization, the KdV equation.
	
\subsubsection{Infinitely many conserved quantities from the Lax operator $L = -\pdr^2 + u$}
\label{s:Infinitely-many-conserved-quantities-for-the-linear-wave-equation}
				
Since $L = -\pdr^2 + u(x,t)$ and $A = c \pdr$ are unbounded differential operators, we do not try to make sense of $\tr L^n$ to find conserved quantities by the method of \S \ref{s:Isospectral-evolution-of-the-Lax-matrix}. Nevertheless, conserved quantities {\it can} be obtained from the pair of equations $L \psi = \la \psi$ and $\psi_t = - A \psi$ (see \S \ref{s:Isospectral-evolution-of-the-Lax-matrix}). Indeed, suppose we put $\la = k^2$ and change variables from the wavefunction $\psi$ to a new function $\rho$ defined via the transformation
	\beq
	\psi(x,t;k)  = \exp\left[- i k x + \int^x_{-\infty} \rho(y,k,t) \: dy \right].
	\label{e:WKB-like-ansatz}
	\eeq 
Then, by studying the quantum mechanical scattering problem for a plane wave with one dimensional wavevector $k$ in the potential $u$ (assumed to vanish at $\pm \infty$), it can be shown \cite{Dunajski} that $\int_{-\infty}^{\infty} \rho(x,k,t) \: dx$ (which is the reciprocal of the transmission amplitude) is conserved in time for any $k$. We will use this to find an infinite sequence of integrals of motion (in terms of $u$). Putting (\ref{e:WKB-like-ansatz}) in $L \psi = k^2 \psi$, we get a Riccati-like equation relating $\rho$ to $u$: $\rho_x + \rho^2 - 2 i k \rho = u(x,t)$. Since $\rho$ is a conserved density, so are the coefficients $\rho_n$ in an asymptotic series in inverse powers of $k$: $\rho = \sum_{n = 1}^{\infty} \rho_n(x,t)/(2 ik)^n$ which is a bit like a semiclassical expansion. Comparing coefficients of different powers of $k$, one finds
	\small
	\beq
	\text{at}  \quad {\cal O}(k^0): \rho_1 = -u, \quad 
	\text{at} \quad {\cal O}(1/k): \rho_2 = \pdr \rho_1 \quad \text{and} \quad
	\text{at}  \quad {\cal O}(1/k^n): \rho_{n+1} = \pdr \rho_n + \sum_{m =1}^{n-1} \rho_m \rho_{n-m}.
	\eeq
	\normalsize
Using this recursion relation we may express $\rho_n$ in terms of $u$ and its derivatives:
	\beqs
	\rho_1 &=& -u, \quad 
	\rho_2 = -u_x, \quad 
	\rho_3 = u^2 - u_{xx}, \quad 
	\rho_4 = (2 u^2 - u_{xx})_x, \cr
	\rho_5 &=& -u_{4x} + 2(u^2)_{xx} + u_x^2 + 2 u u_{xx} - 2 u^3, \quad 
	\rho_6 = \left( -u_{4x} +  18 u u_{2x} - \frac{16}{3} u^3 \right)_x \quad  \text{etc.} 
	\eeqs
The even coefficients integrate to zero while $\rho_{2n+1}$ lead to nontrivial conserved quantities defined as
	\beq
	Q_n = \frac{(-1)^{n+1}}{2} \int_{-\infty}^{\infty} \rho_{2n+1} \: dx \quad \text{for} \quad n = 0,1,2 \ldots.
	\eeq
The first few of these conserved quantities for the wave equation are:
\small
	\beq
	Q_0 = \int \frac{u}{2} \: dx, \quad 
	Q_1 = \int \frac{u^2}{2} \: dx, \quad 
	Q_2 = \int \left( \frac{u_x^2}{2} + u^3 \right) \: dx \quad \text{and} \quad
	Q_3 = \half \int \left[ 5 u^4 + 10 u u_x^2 + u_{2x}^2 \right] \: dx.
	\label{e:conserved-qtys-wave-eqn}
	\eeq	
\normalsize		
	

\subsection{Korteweg-de Vries (KdV) equation}
\label{s:KdV-equation}

\begin{figure}[h]
	\begin{center}
	\includegraphics[width=8cm]{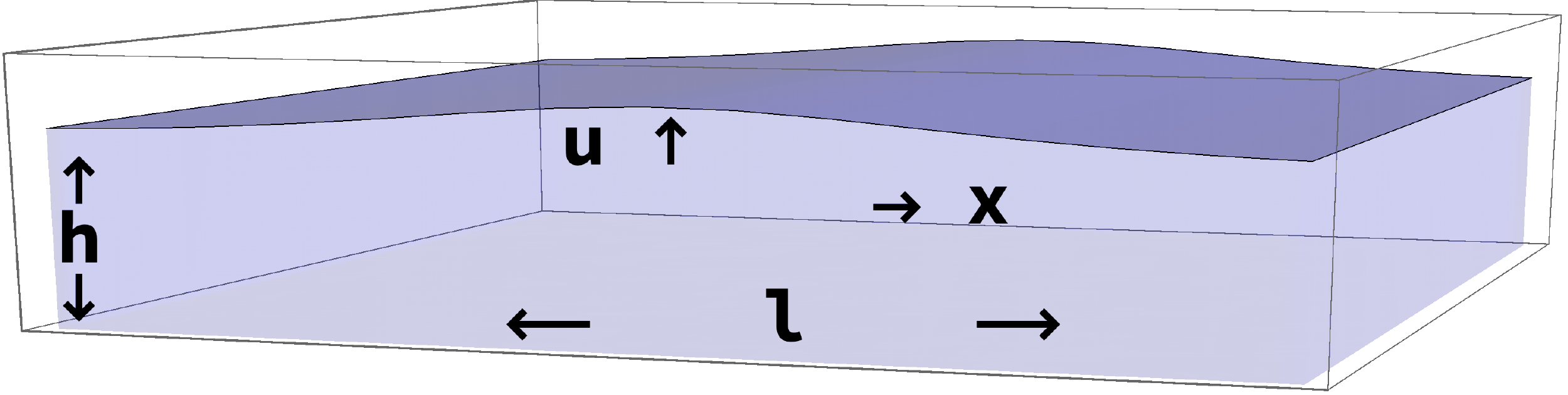}
	\caption{\footnotesize Surface wave profile in a canal.}
	\label{f:KdV-Soliton}
	\end{center}
	\end{figure}

The KdV equation, with subscripts denoting partial derivatives,
	\beq
	u_t - 6 u u_x + u_{xxx} = 0,
	\label{e:KdV-equation}
	\eeq
describes long wavelength ($l \gg h$, `shallow-water') surface waves of elevation\footnote{The standard form (\ref{e:KdV-equation}) of the KdV equation only admits waves of depression as its solutions. To get waves of elevation which we see in a canal, we need to change the sign of the advection term so that the KdV equation takes the form $u_t + 6 u u _x + u_{3x} = 0$.} $u(x,t) \ll h$ in water flowing in a narrow canal of depth $h$ (see Fig.~\ref{f:KdV-Soliton}). 
The KdV equation for the field $u$ describes the evolution of infinitely many degrees of freedom labeled by points $x$ lengthwise along the canal. While the nonlinear advection term $u u_x$ can steepen the slope of a wave profile, the dispersive\footnote{A linear evolutionary partial differential equation (such as the wave equation) is nondispersive if the phase velocity $v_p (k) = \om(k)/k$ of a plane wave solution $e^{i(kx-\om(k) t)}$ is independent of the wavevector $k$. This happens if the angular frequency-wavevector dispersion relation $\om = \om(k)$ is linear. For a nondispersive equation, all Fourier components (labelled by $k$) travel at the same speed so that a wave packet does {\it not} spread out.} $u_{xxx}$  term tends to spread the wave out. A balance between the two effects can lead to localized solitary waves\footnote{The most famous solution of the KdV equation is the {\it soliton} $u = -\frac{c}{2} {\rm sech}^2 \left[ \frac{\sqrt{c}(x-ct)}{2} \right]$. It describes a localized solitary wave of depression that travels at velocity $c$ while retaining its shape. Observation of such a wave was reported in 1834 by Scott Russell while riding along the Edinburgh-Glasgow canal.}  or `solitons' that can propagate while maintaining their shape. What is more, two such solitons can collide and reemerge while retaining their shapes. These phenomena, which were discovered via laboratory and numerical experiments, suggested that the KdV equation may possess several constants of motion.

In fact, the KdV equation admits some elementary conserved quantities \cite{D-J, Das}. For instance, integrating (\ref{e:KdV-equation}) gives
	\beq
	\DD{}{t}\int_{-\infty}^{\infty} u \: dx = \int_{-\infty}^{\infty} \left( 3 u^2 - u_{xx} \right)_x \: dx = 0,
	\label{e:conservation-of-mass-KdV}
	\eeq
assuming $u \to 0$ as $x \to \pm \infty$. This leads to the conservation of the mean height $2 Q_0 = \int_{-\infty}^{\infty} u \: dx$. Furthermore, one may check by differentiating in time and using (\ref{e:KdV-equation}) that 
	\beq
	2 Q_1 = P = \int_{-\infty}^{\infty} u^2 dx \quad \text{and} \quad
	Q_2 = E = \int_{-\infty}^{\infty} \left( u^3 + \frac{u_x^2}{2} \right) dx
	\eeq
are also conserved. $P$ and $E$ can be interpreted as the momentum and energy of the wave and are related to symmetries of the KdV equation under space-time translations via Noether's theorem (see Chapt.~1 of \cite{M-J-D} for more on symmetries of the KdV equation). While these conservation laws could perhaps be guessed, in what came as a major surprise, in 1967-68, Whitham and then Kruskal and Zabusky discovered a fourth ($Q_3$ from (\ref{e:conserved-qtys-wave-eqn})) and fifth conserved quantity. Miura discovered yet more and the list grew to eleven conserved quantities. In fact, it was shown by Gardner, Kruskal and Miura \cite{M-G-K} that the KdV equation admits an infinite sequence of independent conserved quantities\footnote{Interestingly, their work was motivated by Kruskal and Zabusky's 1965 observation \cite{Z-K} of `recurrent behavior' and `soliton scattering' in numerical solutions of the KdV equation.}. They turn out to be the same as the $Q_n$ of \S \ref{s:Infinitely-many-conserved-quantities-for-the-linear-wave-equation} for reasons that will soon become apparent.
 
At around the same time, in 1968, Peter Lax \cite{Lax} proposed an $(L, A)$ pair for the KdV equation. As for the wave equation in \S \ref{s:wave-equation},  $L$ is the Schr\"odinger operator, but $A$ is a third order operator (see Appendix \ref{a:Arriving-at-the-third-order-Lax-operator-for-the-KdV-equation} for an indication of how one arrives at $A$):
	\beq
	L = -\pdr^2 + u(x, t) \quad \text{and} \quad 
	A = 4 \pdr^3 - 6 u \pdr - 3 u_x.
	\label{e:lax-pair-KdV}
	\eeq
As before, $L_t = u_t$. The commutator $[L,A]$ receives two contributions. With $u'$ denoting $u_x$, the $3^{\rm rd}$ order term in $A$ gives
	\beq
	[-\pdr^2 + u, 4 \pdr^3] = -4(u''' + 3u'' \pdr + 3 u' \pdr^2).
	\eeq
As for the first order part of $A$, the calculation is essentially the same as in (\ref{e:Wave-eqn-Lax-pair-commutator}) of Appendix \ref{a:Finding-a-Lax-operator-$A$-for-the-wave-equation}, with $\al = -3 u$:
	\beq
	[-\pdr^2 + u, -3(2 u \pdr + u_x)] = -3(u''' - 4 u'' \pdr - 4 u' \pdr^2 - 2 u u').
	\eeq
Adding these, the differential operator terms in $L_t = [L, A]$ cancel, leaving us with the KdV equation (\ref{e:KdV-equation}):
	\beq
	u_t = [L, 4 \pdr^3 - 6 u \pdr - 3 u_x] = -u''' + 6 u u'.
	\eeq
The Lax representation helps us understand roughly why KdV admits infinitely many conserved quantities. Indeed, $L = - \pdr^2 + u$ may be viewed as an infinite dimensional matrix, all of whose eigenvalues are conserved.  In fact, the method of \S \ref{s:Infinitely-many-conserved-quantities-for-the-linear-wave-equation} for finding conserved quantities for the wave equation from its Lax operator $L$ also applies to the KdV equation. What is more, since the two equations share the same Lax operator $L$, it turns out that they also possess the same set of conserved quantities $Q_n$. Moreover, treating $Q_n$ as a sequence of  Hamiltonians, one obtains the `KdV' hierarchy of field equations. The linear wave and KdV equations are the first two in this hierarchy, while $u_t = u_{5x} -10 u u_{3x} - 20 u_x u_{2x} + 30 u^2 u_x$ is the third. The Schr\"odinger operator $L = -\pdr^2 + u$ serves as a common Lax operator for all of them though the operator $A$ (which enters through $\psi_t = -A \psi$) differs for the various members of this hierarchy. Remarkably, it turns out that the $Q_n$ of \S \ref{s:Infinitely-many-conserved-quantities-for-the-linear-wave-equation} are integrals of motion for each of the equations in this hierarchy.


\subsection{From Lax pair to zero curvature representation}
\label{e:From-Lax-pair-to-zero-curvature-representation}

The zero curvature representation generalizes the idea of a Lax pair to a wider class of nonlinear evolution equations for systems especially in one spatial dimension. To understand how this works, we change our viewpoint and regard the nonlinear Lax equation $L_t = [L, A]$ as a compatibility condition for the following pair of linear equations to admit simultaneous solutions:  
	\beq
	L \psi = \la \psi \quad \text{and} \quad 
	\psi_t = - A \psi \quad \text{with} \quad \la \quad \text{a constant}.
	\label{e:two-linear-eqns-scalar-Lax}
	\eeq
Indeed, by differentiating $L \psi = \la \psi$ in time and using the second equation, it is verified that for the  eigenvalue $\la$ of $L$ to be time-independent, $L$ and $A$ must satisfy the Lax equation $L_t = [L,A]$. Unlike in \S \ref{s:Isospectral-evolution-of-the-Lax-matrix}, here there is no need for $\la$ to be a nondegenerate eigenvalue of $L$.

In the case of the KdV equation (\ref{e:KdV-equation}), $L = -\pdr_x^2 + u$ involves $2^{\rm nd}$ order space derivatives, so that the two equations in (\ref{e:two-linear-eqns-scalar-Lax}) are somewhat asymmetrical. There is a way of replacing (\ref{e:two-linear-eqns-scalar-Lax}) with a more symmetric pair of linear equations involving only $1^{\rm st}$ order derivatives:
	\beq
	\pdr_x F = U F \quad \text{and} \quad \pdr_t F = V F.
	\label{e:auxiliary-linear-eqns}
	\eeq
The price to be paid is that $U$ and $V$ are now square matrices and $F$ a column vector (of size equal to the order of the differential operator $L$) whose components depend on location through the dynamical variables (such as $u$ for KdV). The matrix elements of $U$ and $V$ also depend on the eigenvalue $\la$ which is now called the spectral parameter. However, unlike $L$ and $A$ which are differential operators, $U$ and $V$ are finite dimensional matrices, a feature we will exploit in obtaining conserved quantities.

Eqn. (\ref{e:auxiliary-linear-eqns}) is called the auxiliary linear system of equations of Zakharov and Shabat \cite{Z-S-1, Z-S-2}. $\pdr_x - U$ and $\pdr_t - V$ may be viewed as the space and time components of a `covariant derivative'. Thus, the auxiliary linear equations (\ref{e:auxiliary-linear-eqns}) require that every vector field $F(x,t)$ is `covariantly' constant. It is overdetermined in the sense that $U$ and $V$ must satisfy a compatibility (consistency) condition for solutions $F$ to exist. Indeed, equating mixed partials $\pdr_x \pdr_t F = \pdr_t \pdr_x F$, we get the consistency condition 
	\beq
	\pdr_t U - \pdr_x V + [U, V] = 0.
	\label{e:zero-curvature-condition}
	\eeq
The original nonlinear evolution equations are said to have a zero curvature representation if they are equivalent to (\ref{e:zero-curvature-condition}) for some pair of matrices $U$ and $V$. Before explaining how this scheme may be used to find conserved quantities, let us use the KdV equation to provide an example.

To find $U$ for KdV, we write the eigenvalue problem for the Lax operator $(-\pdr_x^2 + u) \psi = \la \psi$ as a pair of first order equations\footnote{For a Lax operator which is an $n^{\rm th}$-order spatial differential operator, we may express the Lax equation as a system of $n$ first order equations for the column vector $(\psi, \psi_x, \psi_{xx} , \cdots, \psi_{(n-1)x})^T$ comprising the first $(n-1)$ derivatives of the eigenfunction $\psi$. $U$ and $V$ then become $n \times n$ matrices. For the KdV equation, $n = 2$.} by introducing the column vector $F = (f_0, f_1)^T = (\psi, \psi_x)^T$: 
	\beq
	\pdr_x \colvec{2}{f_0}{f_1} = \colvec{2}{0 & 1}{u-\la & 0} \colvec{2}{f_0}{f_1} \quad \imply \quad
	U = \colvec{2}{0 & 1}{u-\la & 0},
	\eeq
upon comparing with (\ref{e:auxiliary-linear-eqns}). Next, we use $\psi_t = - A \psi$ with $A \psi = 4 \psi_{xxx} - 6 u \psi_x - 3 u_x \psi$ to find $V$ such that $\pdr_t F = V F$. We may express $\psi_t = - A \psi$ as a system of two first order ODEs. First, we differentiate $L \psi = \la \psi$ in $x$ to express $\psi_{xxx}$~ as ~$u_x \psi + u \psi_x - \la \psi_x$. Thus, $A \psi$ can be written in terms of $\psi$ and $\psi_x$:
	\beq
	A \psi = -2(u + 2 \la ) \psi_x + u_x \psi.
	\eeq
Next, using $F = (\psi, \psi_x)^T$, $\psi_t = - A \psi$ takes the form
	\beqs
	\pdr_t \colvec{2}{f_0}{f_1} =V \colvec{2}{f_0}{f_1} \quad
	\text{with} \quad V = \colvec{2}{-u_x & 2(u + 2 \la)}{2u^2 - u_{xx} + 2u \la - 4 \la^2 & u_x}.
	\label{e:V-for-KdV}
	\eeqs 
Here, the second row of the matrix $V$ is obtained by taking the $x$ derivative of the first equation in (\ref{e:V-for-KdV}) and using $L \psi = \la \psi$. The parameter $\la$ that appears in $U$ and $V$ originally arose as the eigenvalue of the Lax operator $L$. This explains the name {\it spectral parameter}. More generally, a zero curvature representation need not arise from a Lax pair and the corresponding spectral parameter $\la$ may not admit an interpretation as an eigenvalue.

\vspace{.5cm}


{\fl \bf Why the name `zero curvature'?} In general relativity the gravitational field is associated to space-time curvature. It turns out that an electromagnetic field is {\it also} associated to curvature, though not of space-time but of an internal space (U(1) principal bundle over space-time). Now, the electric and magnetic fields may be packaged in the components of the field strength: $F_{0i} = E_i/c$ and $F_{ij} = \sum_k \eps_{ijk} B_k$ for $1 \leq i,j,k \leq 3$, where $c$ denotes the speed of light. Thus, the field strength is a measure of curvature. What is more, specializing to one spatial dimension ($x^0 = t, x^1 = x$) and introducing the scalar and vector potentials $A_0$ and $A_1$, we have $F_{01} = \pdr_t A_1 - \pdr_x A_0$. More generally, in the non-abelian gauge theories relevant to the strong and weak interactions, $A_0$ and $A_1$ become square matrices and the field strength acquires an extra commutator term: $F_{01} =  \pdr_t A_1 - \pdr_x A_0 + [A_1, A_0]$. Now making the substitutions $A_1 \to U$ and $A_0 \to V$, we see that the consistency condition (\ref{e:zero-curvature-condition}) states that the field strength or curvature of this nonabelian gauge field vanishes. Hence the name zero curvature condition.
	

\subsection{Conserved quantities from the zero curvature condition}
\label{s:Conserved-quantities-from-the-zero-curvature-condition}

Here, we will learn how the zero curvature representation may be used to construct conserved quantities. Let us consider the first of the auxiliary linear equations in (\ref{e:auxiliary-linear-eqns}) for the column vector $F$: $\pdr_x F = U(x) F(x)$. Let us imagine solving this equation for $F$ from an initial location $x$ to a final point $y$. If $y = x + \del x$ for small $\del x$, then
	\beq
	F(x + \del x) \approx \left[{\bf 1} + \del x \: U(x)\right] \: F(x). 
	\label{e:infintesimal-evolution-F}
	\eeq
More generally, linearity suggests that the solution may be written as $F(y) = T(y, x) F(x)$. Here $T(y,x)$ may be viewed as transforming $F(x)$ into $F(y)$ and is called the {\it transition matrix} or sometimes a parallel transport operator. For this to work, $T(y,x)$ must satisfy the equation and boundary condition 
	\beq
	\pdr_y T(y, x; \la) = U(y; \la) T(y, x; \la) \quad \text{and} \quad T(x, x; \la) = {\bf 1}
	\label{e:Transition-derivative}
	\eeq
for any value of the spectral parameter $\la$. This is obtained by inserting $F(y) = T(y, x) F(x)$ in the auxiliary linear equation $\pdr_y F(y) = U(y) F(y)$ and requiring it to hold for any $F(x)$. In Appendix \ref{a:Time-evolution-operator-and-the-ordered-exponential}, we learn that the transition matrix $T(y,x)$ may be expressed (essentially by iterating (\ref{e:infintesimal-evolution-F})) as an {\it ordered exponential series} which we abbreviate as 
	\beq
	T(y, x; \la) = {\rm P}\exp \int_{x}^y U(z; \la) \: dz.
	\label{e:transition-matrix}
	\eeq
For simplicity, we henceforth suppose that our one-dimensional system is defined on the spatial interval $-a \leq x \leq a$ with {\it periodic boundary conditions}, so that $U(-a) = U(a)$ and $V(-a) = V(a)$. Thus, we may view our spatial coordinate $x$ as parametrizing a circle of circumference $2a$. So far, we have been working at one instant of time. It turns out that the transition matrix around the full circle ($x = -a$ to $y = a$), which is also called the monodromy matrix,
	\beq
	T_a(t, \la) = {\rm P}\exp \int_{-a}^a U(z; t, \la) \: dz
	\eeq
has remarkably simple time evolution. In fact, using the derivative of the transition matrix (\ref{e:Transition-derivative}) and the zero curvature condition (\ref{e:zero-curvature-condition}), one may show (see Appendix \ref{a:Time-evolution-operator-and-the-ordered-exponential} or \S 3 of Chapter 1 of \cite{F-T}) that the transition matrix evolves according to:
	\beq
	\pdr_t T(y, x; t) = V(y; t) T(y, x; t) - T(y, x; t) V(x; t).
	\label{e:evolution-of-transition-matrix}
	\eeq
However, this is not quite a commutator. Nevertheless, specializing to $x = -a$ and $y = a$ and using periodic boundary conditions, we find that the monodromy matrix $T_a(t) = T(a,-a; t)$ evolves via a commutator: 
	\beq
	\pdr_t T_a(t, \la) = [V(a; t, \la), T_a(t, \la)].
	\label{e:monodromy-time-evolution}
	\eeq
We are now in familiar territory: this equation has the same structure as the Lax equation (\ref{e:Lax-equation}) upon making the replacements $T_a \mapsto L$ and $V \mapsto -A$. As explained in \S \ref{s:Isospectral-evolution-of-the-Lax-matrix}, the spectrum of the Lax matrix $L$ is independent of time. This immediately implies that the trace of the monodromy $\tr T_a(t, \la)$ is independent of time\footnote{This is where the finite dimensional character of $U$,$V$ and consequently the monodrmy matrix $T_a$ is useful. Unlike the trace of the Lax differential operator $L$, there is no difficulty in making sense of the trace of the monodromy matrix.}. Moreover, this is true for any value of the spectral parameter $\la$. Thus, if we expand $\tr T_a(\la)$ in a series in (positive and negative) powers of $\la$, then each of the coefficients is a conserved quantity. In many interesting cases such as the Heisenberg magnetic chain and nonlinear Schr\"odinger equations, one may obtain infinitely many conserved quantities in this way. 

\subsection{Nonlinear Schr\"odinger equation}
\label{s:NLSE}

We now briefly illustrate the idea of a zero curvature representation by considering the one-dimensional nonlinear Schr\"odinger equation (NLSE) for the complex wave amplitude $\psi(x,t)$:
	\beq
	i \dd{\psi}{t} = -\frac{\pdr^2 \psi}{\pdr x^2} + 2 \kappa |\psi|^2 \psi.
	\label{e:NLSE}
	\eeq
Here, $\kappa$ is a real parameter. For $\kappa =0$, it reduces to the linear Schr\"odinger wave equation (in units where $\hbar = 1$) for a free quantum mechanical particle of mass $m = 1/2$ moving on a line. The NLSE is used to model a gas of bosons with short-range pairwise interactions (of strength $\kappa$) in a `mean field' approximation where $|\psi|^2$  is interpreted as the density of bosons. It also has applications in nonlinear optics \cite{A-S}. Like KdV, NLSE too admits solitary wave solutions. They are called bright and dark solitons depending on whether $\kappa$ is negative or positive corresponding to attractive or repulsive interactions among the bosons.

The NLSE admits a zero curvature representation (\ref{e:zero-curvature-condition}) if the $U$ and $V$ matrices are chosen as \cite{F-T} where $\la$ is an arbitrary spectral parameter:
	\beqs
	U &=& U_0 + \la U_1 \quad \text{and} \quad
	V = V_0 + \la V_1 + \la^2 V_2 \quad \text{where} \quad
	U_0 = \sqrt{\kappa} (\psi^* \sigma_+ + \psi \sigma_-) = -V_1, \cr
	U_1 &=& \frac{\sigma_3}{2i} = -V_2 \quad \text{and} \quad
	V_0 = i \kappa |\psi|^2 \sigma_3 - i \sqrt{\kappa} \left(\psi^*_x \sigma_+ - \psi_x \sigma_- \right).
	\eeqs
Here, $\sig_{\pm} = (1/2)(\sig_1 \pm i \sig_2)$ are built from the Pauli matrices. It may be checked that the condition for the associated field strength (\ref{e:zero-curvature-condition}) to vanish for all values $\la$ is equivalent to the NLSE and its complex conjugate. Thus, $(V, U)$ may be viewed as defining a flat connection in an SU(2) principal bundle over the 1+1-dimensional space-time. 

Using the methods of \S \ref{s:Conserved-quantities-from-the-zero-curvature-condition}, with some more effort, it is possible to find an infinite sequence of conserved quantities for the NLSE. For bosons moving in the interval $-a \leq x \leq a$ with periodic boundary conditions, the first four integrals of motion are 	
	\beqs
	N &=& \int_{-a}^a |\psi|^2 \, dx, \quad
	P = \int_{-a}^a \Im \psi^* \psi_x \, dx, \quad
	E = \int_{-a}^a (|\psi_x|^2 + \kappa |\psi|^4) \, dx \cr
	\text{and} \quad
	Q &=& \int_{-a}^a \left[ \psi^* \psi_{xxx} - \kappa |\psi|^2 \left( \psi \psi^*_x + 4 \psi^* \psi_x \right) \right] \, dx.
	\eeqs
The conserved quantities $N, P$ and $E$ represent the number of bosons, their total momentum and energy.

\section{Discussion}
\label{s:Epilogue}

In this article, we have explained what a Lax pair is and how it can be used to find conserved quantities for mechanical systems such as the simple harmonic oscillator, Toda chain, Euler top and Rajeev-Ranken model as well as field theories such as the linear wave and KdV equations. Though it is not always possible or easy to find a Lax pair for a given system, it is possible to generate lots of Lax pairs and thereby discover systems with numerous conserved quantities. Some of these turn out to be interesting `exactly solvable' or `integrable' systems. As one may infer from these examples, there is no step-by-step procedure to find a Lax pair for a given system or even to know whether it admits a Lax pair. One first needs to determine some properties of the system (say numerically, analytically or experimentally as happened with KdV) to develop a feeling for whether a Lax pair might exist. As a rule of thumb, equations whose trajectories are `regular' or for which (some) analytic solutions can be obtained often do admit a Lax pair, while those that display irregular/chaotic behavior do not. Even if one suspects the presence of a Lax pair, finding one may not be easy and requires playing around with the equations as we have done for the harmonic oscillator, Euler top, wave equation and the KdV equation. However, if one {\it does} find a Lax pair, it opens up a whole new window to the problem and brings to bear new tools \cite{GGKM} that can be applied to its understanding. Indeed, Lax pairs are the tip of an iceberg in the study of (Hamiltonian) dynamical systems. While it helps to have conserved quantities, one can do more if they are sufficiently numerous and generate `commuting' flows on the state space (i.e., if their Poisson brackets vanish). In such cases, there is (at least in principle) a way of changing variables to so-called action-angle variables in which the solutions to the EOM may be written down by inspection! Moreover, continuum systems in {\it one spatial dimension} (such as the KdV, nonlinear Schr\"odinger and sine-Gordon equations) which have a Lax pair and an infinite tower of conserved quantities typically admit solitary wave solutions called solitons. Two such solitons can collide with each other and interact in a complicated way but emerge after the collision retaining their original shapes and speeds, thus mimicking the elastic scattering of particles. This {\it soliton scattering} behavior can be regarded as a generalization to nonlinear systems of the superposition principle for linear equations. These nonlinear field equations also admit a remarkable generalization of the Fourier transform technique of solving linear PDEs such as the heat or wave equations. This technique is based on the Gelfand-Levitan-Marchenko equation and is called the `inverse scattering transform' \cite{D-J, F-T, A-S, N-M-P-Z}: it can be used to solve the initial value problem of determining the fields at time $t$ given their values at $t = 0$.


\appendix

\section{Finding a Lax operator $A$ for the wave equation}
\label{a:Finding-a-Lax-operator-$A$-for-the-wave-equation}
			
The choice $A = c \pdr$ to partner the Schr\"odinger operator $L$ in the Lax pair  (\ref{e:wave-eqn-Lax-eqn-form}) for the wave equation (\ref{e:1D-wave-eqn}) can be arrived at by starting from the simplest of differential operators, a first order differential operator $\al (x, t) \pdr + \beta (x, t)$ and imposing some consistency conditions. We shall see in Appendix \ref{a:Arriving-at-the-third-order-Lax-operator-for-the-KdV-equation}, that this approach generalizes to other equations. To make $A$ anti-symmetric, we subtract its adjoint and consider
	\beq
	A_1 = (\al \pdr + \beta - \pdr^{\dagger} \al - \beta) = ( \al \pdr + \pdr \al ) = [\al, \pdr]_+ = (\al' + 2 \al \pdr) \quad \text{where} \quad \al' = \dd{\al}{x}.
	\label{e:1st-order-diff-anti-symm}
	\eeq
Here, we used (i) $\pdr^{\dagger} = - \pdr$, (ii) $g^{\dagger} = g$ for any real function $g$ and (iii) $(\pdr \al)(f) = \al' f + \al f'$ so that $\pdr \al = \al' + \al \pdr$. The commutator with the Schr\"odinger operator $L$ is then
	\beq
	[L, A_1] = [-\pdr^2 + u , \al' + 2 \al \pdr] = -\al''' - 4 \al'' \pdr - 4 \al' \pdr^2 - 2 \al u'.
	\label{e:Wave-eqn-Lax-pair-commutator}
	\eeq
Here, we used $[\pdr, \al] = \al'$, the Leibnitz product rule, linearity and anti-symmetry of commutators to obtain
	\beqs
	[u, \al'] &=& 0, \quad [u, 2 \al \pdr] = -2 \al u', \quad [-\pdr^2, \al'] = -\pdr[\pdr , \al'] - [\pdr, \al'] \pdr =  -\al''' - 2 \al'' \pdr \quad\text{and} \cr
	[-\pdr^2, 2 \al \pdr] &=& -\pdr[\pdr , 2 \al \pdr] - [\pdr, 2 \al \pdr] \pdr = -\pdr(2 \al' \pdr) - (2 \al' \pdr) \pdr = -4 \al' \pdr^2 - 2 \al'' \pdr.
	\eeqs
In the Lax equation $L_t = [L, A_1]$, $L_t = u_t$ is multiplication by $u_t(x, t)$. For $[L, A_1]$ in (\ref{e:Wave-eqn-Lax-pair-commutator}) to also be a multiplication operator, the coefficients of $\pdr$ and $\pdr^2$ must vanish which implies $\al' = \al'' \equiv 0$ for all $x$. This implies $\al = \al(t)$ is a function of time alone. Thus, $L_t = [L, A_1]$ becomes $u_t = -2 \al(t) u_x$.  For this to be equivalent to the wave equation $u_t + c u_x = 0$, we must pick $\al(t) = c/2$, so that $A_1$ reduces to $ A = c \pdr$.

\section{Arriving at the $3^{\rm rd}$ order Lax operator $A$ for KdV}
\label{a:Arriving-at-the-third-order-Lax-operator-for-the-KdV-equation}

Here we adapt the method of Appendix \ref{a:Finding-a-Lax-operator-$A$-for-the-wave-equation} to explain the choice of the 3$^{\rm rd}$ order differential operator 
	\beq
	A = 4 \pdr^3 - 6 u \pdr - 3 u_x = 4 \pdr^3 - 3 [u, \pdr]_+
	\eeq
in the KdV Lax pair (\ref{e:lax-pair-KdV}). From \S \ref{s:wave-equation}, we know that $A = c \pdr$ and the Schr\"odinger operator $L = -\pdr^2 + u$ furnish a Lax pair for the linear wave equation. To find a Lax pair for the 3$^{\rm rd}$ order KdV equation, we will retain $L = -\pdr^2 + u$ with $L_t$ being the multiplication operator $u_t$, while allowing for $A$ to be of order higher than one. The simplest possibility is a 2$^{\rm nd}$ order operator, but this does not work. Indeed, anti-symmetrization reduces it to a 1$^{\rm st}$ order operator which is no different from (\ref{e:1st-order-diff-anti-symm}) with $\al = -(e' + g f')$:
	\beq
	A_2 = e \pdr^2 + f \pdr g \pdr - (e \pdr^2 + f \pdr g \pdr)^{\dagger} 
	= e \pdr^2 + f \pdr g \pdr - \pdr^2 e - \pdr g \pdr f
	= -(e' + g f')' - 2(e' + g f') \pdr.
	\eeq  
 The next possibility is a 3$^{\rm rd}$ order operator. For simplicity, we try the operator $b \pdr^3$ where $b$ is a constant. Upon anti-symmetrizing,
	\beq
	A_3 = b \pdr^3 - (b \pdr^3)^{\dagger} = b \pdr^3 + \pdr^3 b = 2 b\pdr^3.
	\eeq
As in Appendix \ref{a:Finding-a-Lax-operator-$A$-for-the-wave-equation}, using the product rule and $[\pdr, h] = h'$ we find that
	\beq
	[L, A_3] = [-\pdr^2 + u, 2 b\pdr^3] = -2b(u''' + 3 u'' \pdr +3 u' \pdr^2).
	\eeq
While this includes a $u'''$ term, it lacks the $u u'$ term in the KdV equation (\ref{e:KdV-equation}) and is not purely a multiplication operator. Here, $A_1$ from  Appendix \ref{a:Finding-a-Lax-operator-$A$-for-the-wave-equation} comes to the rescue. Thus, let us consider $A = A_3 + A_1 = 2 b \pdr^3 + 2 \al \pdr + \al'$ so that
	\beq
	[L, A] = [-\pdr^2 + u, 2 b \pdr^3 + 2 \al \pdr + \al' ] = -\al''' - 2 \al u' -2 b u''' - (6 b u''+ 4 \al'') \pdr - (6 b u' + 4 \al') \pdr^2.
	\eeq 
For $[L,A]$ to be a multiplication operator, the coefficients of $\pdr$ and $\pdr^2$ must vanish. Thus, $\al' = -(3/2) b u'$ which implies $ \al = -(3/2) b u + \al_0$ for an integration constant $\al_0$. Eliminating $\al$, the Lax equation becomes
	\beq
	L_t = u_t = [L, A] = -\frac{b}{2}u''' + (3 b u - 2 \al_0) u'.
	\eeq
Comparing with the KdV equation ($u_t = 6 u u_x - u_{3x}$) fixes $b = 2$ and $\al_0 = 0$ so that $A = 4 \pdr^3 - 6 u \pdr - 3 u'$ as claimed. Note that we may add to $A$ an arbitrary function of time (which would commute with $L$) without affecting the Lax equation.	


\section{Time evolution operator and the ordered exponential}
\label{a:Time-evolution-operator-and-the-ordered-exponential}

In \S \ref{s:Lax-pairs-for-mechanical-systems}, we encountered an equation for the time evolution operator $S(t)$
	\beq
	\dot S = -A(t) \, S, \quad \text{with the initial condition} \quad S(0) = {\bf 1},  \quad \text{the identity matrix}.
	\label{e:time-evolution-equation}
	\eeq
The same equation also arises as the second of the auxiliary linear equations in (\ref{e:auxiliary-linear-eqns}) and as the Schr\"odinger equation in quantum mechanics for the time dependent `Hamiltonian' $-i \hbar A(t)$. Here, we explain how this equation may be solved. When $A$ is independent of time the solution is the matrix exponential $S = \exp(-At)$. However, for time-dependent $A$, this formula does not satisfy (\ref{e:time-evolution-equation}) if $A(t)$ at distinct times do not commute. To solve (\ref{e:time-evolution-equation}), we first integrate it in time form 0 to $t$ to get an integral equation that automatically encodes the initial condition:
	\beq
	S(t) - {\bf 1} = -\int_{0}^t A(t_1) S(t_1) \: dt_1.
	\eeq
$S$ appears on both sides, so this is not an explicit solution. Iterating once, we get
	\beq
	S(t) = {\bf 1} - \int_{0}^t dt_1 \: A(t_1) \left( {\bf 1} - \int_{0}^{t_1} dt_2 \: A(t_2) S(t_2) \right).
	\eeq
Repeating this process, we get an infinite sum of multiple integrals,
	{\scriptsize
	\beq
	S(t) = {\bf 1} - \int_{0}^t dt_1 \: A(t_1) + \int_{0}^t \int_{0}^{t_1} dt_1  dt_2  \: A(t_1) A(t_2) - \cdots
	       = \sum_{n = 0}^{\infty} (-1)^n \int \cdots \int_{0 < t_n <  \cdots  < t_1 < t} dt_1 \cdots dt_n \: A(t_1) A(t_2) \cdots A(t_n). 	\qquad
	\eeq}
Now, if we define time ordering denoted by the symbol ${\rm T}$ via
	\beq
	{\rm T}(A(t_1) A(t_2)) = \begin{cases}
				    A(t_1) A(t_2)  \quad \text{if} \quad t_1 \geq t_2 \\ 
				    A(t_2) A(t_1) \quad \text{if}  \quad t_2 \geq t_1,
				    \end{cases}
	\label{e:time-ordering}
	\eeq
and use the identity $\int_{t_1 > t_2} dt_1 dt_2 A(t_1) A(t_2)  = \int_{t_2 > t_1} dt_1 dt_2 A(t_2) A(t_1)$, we obtain
	\beq
	\int_{0}^{t} dt_1 \int_0^{t_1} dt_2 A(t_1)A(t_2) = \half \int_0^{t} dt_1 \int_{0}^{t} dt_2 \: {\rm T}(A(t_1) A(t_2)).
	\eeq
Thus, we have expressed an integral over a triangle in the $t_1{\text-}t_2$ plane as half the integral over a square. Similarly, for $n=3$ we may express the integral over a pyramid as one/sixth of that over a cube. Proceeding this way, we get
\footnotesize
	\beqs
	\int_0^t dt_1 \int_0^{t_1} dt_2 && \cdots \int_0^{t_{n-1}} dt_n \; A(t_1) A(t_2) \cdots A(t_n) = \frac{1}{n!} \int_0^t \cdots \int_0^t dt_1 dt_2 \cdots dt_n {\rm T}(A(t_1) A(t_2) \cdots A(t_n)) \cr \text{so that} \;\;
	S(t) &=&  \sum_0^\infty \frac{(-1)^n}{n!} \int_0^t \cdots \int_0^t dt_1 dt_2 \cdots dt_n {\rm T}(A(t_1) A(t_2) \cdots A(t_n)) =: {\rm T}\exp \left[-\int_0^t A(t')dt' \right]. \qquad
	\eeqs
\normalsize
This series is called a time-ordered exponential and denoted ${\rm T}\exp$. If time is replaced with a spatial coordinate, then it is called a path-ordered exponential and abbreviated `${\rm P}\exp$'.

\section{Time evolution of the transition matrix $T(y,x;t)$}
\label{a:Time-evolution-of-the-transition-matrix-$T(y,x;t)}

Recall from \S \ref{s:Conserved-quantities-from-the-zero-curvature-condition} that the transition matrix $T(y,x;t)$ `propagates' vectors in the auxiliary linear space from $x$ to $y$: $F(y;t) = T(y,x;t) F(x;t)$ and may be expressed as a path ordered exponential as in (\ref{e:transition-matrix}). To obtain Eqn. (\ref{e:evolution-of-transition-matrix}) for its time evolution, we first differentiate Eqn. (\ref{e:Transition-derivative}) [$\pdr_y T(y,x;t) = U(y;t) T(y,x;t)$] in time:
	\beq
	\pdr_t \pdr_y \: T(y,x;t) =  \pdr_t U(y;t) \: T (y,x;t) + U(y;t) \: \pdr_t T(y,x;t).
	\eeq
Then we use the zero curvature condition $\pdr_t U(y) -  \pdr_y V(y) + [U(y), V(y)] = 0$ and  Eqn. (\ref{e:Transition-derivative}) again to get:
	\beq
	\pdr_t \pdr_y \: T(y,x;t) = (\pdr_y V) T + V U T - U V T + U (\pdr_t T) = \pdr_y(VT) + U (\pdr_t T - V(y) T).
	\eeq
Thus, we have
	\beq
	\pdr_y W(y,x;t)= U(y) W(y,x;t) \quad \text{where} \quad W(y,x;t) = \pdr_t T - V(y) T.
	\label{e:W-derivative-eqn}
	\eeq
Thus both $W(y,x;t)$ and $T(y,x;t)$ satisfy the {\it same} differential equation (\ref{e:Transition-derivative}) though they obey different `boundary conditions' $W(x,x;t) = -V(x)$ while $T(x,x;t) = I$. We now use this to check that $\tl W(y,x;t) = - T(y,x;t) V(x)$ also satisfies the same differential equation with the desired boundary condition $\tl W(x,x;t) = -V(x)$. Exploiting the uniqueness of solutions to (\ref{e:Transition-derivative}) for a given boundary condition, we conclude that $W (y,x;t) = \tl W = - T(y,x;t) V(x)$. Substituting this in the definition of $W(y,x;t)$ (\ref{e:W-derivative-eqn}), we obtain the evolution equation  (\ref{e:evolution-of-transition-matrix}) for the transition matrix: $\pdr_t T(y, x; t) = V(y; t) T(y, x; t) - T(y, x; t) V(x;t)$.

\vspace{.25cm}

{\fl \bf Acknowledgements:} We thank an anonymous referee for useful comments and references. This work was supported in part by the Infosys Foundation, J N Tata Trust and grants (MTR/2018/000734, CRG/2018/002040) from the Science and Engineering Research Board, Govt. of India.



\end{document}